\documentclass[preprint,review,12pt]{elsarticle}

\journal{Applied Surface Science}

\usepackage[margin=1.5cm,bottom=1.5cm]{geometry}

\newcommand{\hl}[1]{#1}
\usepackage{amssymb}
\usepackage{amsmath}
\usepackage{graphicx}
\usepackage{dcolumn}
\usepackage{bm}
\usepackage{etoolbox}
\graphicspath{{final_figs/}{figs/}}
\usepackage[colorlinks=true, allcolors=blue]{hyperref}
\usepackage{xcolor}
\usepackage{subcaption}
\usepackage{caption}
\usepackage{multirow}
\usepackage{gensymb, xspace, verbatim}
\usepackage[version=4]{mhchem}

\usepackage{afterpage}     
\usepackage{placeins}

\newcommand{\HeMin}{0\,\%\,He\xspace}
\newcommand{\HeMax}{82\,\%\,He\xspace}

\newcommand{\li}[1]{\textsubscript{#1}}
\newcommand{\latin}[1]{{\em{}#1}}

\begin{document}

\begin{frontmatter}
\title{Enhanced hydrogen response of copper-doped TiO\li{2} synthesised by helium-assisted magnetron sputtering} 

\author[A1]{Akash Kumar}
\author[A1]{Nirmal Kumar}
\author[A1]{David Kolenat\'y}
\author[A1]{Mina Farahani}
\author[A1]{Ji\v{r}í Rezek}
\author[A1]{Tom\'a\v{s} Koz\'ak}
\author[A1,cor1]{Stanislav Haviar}  
\cortext[cor1]{haviar@fav.zcu.cz}    

\affiliation[A1]{organization={Department of Physics and NTIS -- European Centre of Excellence, University of West Bohemia in Pilsen},
              addressline={Univerzitní~8}, 
              city={Pilsen},
              postcode={301 00}, 
              country={Czech Republic}}

\begin{abstract}
Cu-doped TiO$_2$ thin films for hydrogen sensing were synthesised by reactive DC magnetron sputtering in Ar/O$_2$/He mixtures, with the He fraction used as a control parameter for film growth. By combining normal-angle deposition (NAD) and glancing-angle deposition (GLAD) with post-deposition annealing, the effects of He on microstructure formation and sensor performance were examined. X-ray diffraction and electron microscopy revealed that He promotes nanostructuring, lattice expansion in as-deposited NAD films, increased porosity after annealing, and a stronger anatase character in the final oxide layers. These structural changes, which enhance the reactive surface area, lead to improved hydrogen sensing at 300\,$^\circ$C in 1~vol.\,\% H$_2$. The response of NAD films increased from 1.4 to 6.0 simply by replacing part of the argon with helium, whereas GLAD films showed only a modest increase. The observed nanostructuring is discussed in terms of a simulation-supported growth scenario involving energetic backscattered He, a reduced hammering effect, and cooling-related suppression of adatom mobility, which together favour the formation of a more open sensing layer. Helium-assisted sputtering represents a useful physical route for tailoring oxide thin films for gas-sensing applications.
\end{abstract}


\end{frontmatter}
\section{Introduction}

Hydrogen is recognised as a clean, abundant energy source with significant potential for sustainable energy systems. It powers fuel cells for zero-emission vehicles, facilitates renewable energy storage \cite{Schlapbach2001353}, and is essential in chemical production and metal processing industries. As a versatile energy source, hydrogen supports environmental sustainability by enabling low-carbon solutions in transportation, industrial processes, and power generation. However, hydrogen's high combustibility and wide explosive range in a mixture with air (from 4\,\%) pose substantial safety risks \cite{HRUSKA2024, Dagdougui2018}, necessitating robust detection and monitoring systems. There are various gas-sensing technologies, such as electrochemical \cite{Korotcenkov2009}, conductometric metal oxide semiconductor-based (MOS) \cite{Patel1999}, optical \cite{Kulikova2023}, capacitance-based \cite{Huimin2021}, and others \cite{ Gautam2024}. Among various detection technologies, conductometric sensors based on metal oxide semiconductors are prominent for their stability and low cost \cite{Moseley2017}. Among various metal oxides, \ce{SnO2} \cite{Gautam2024, PARK2024}, \ce{TiO2} \cite{Li2018, HAIDRY2020144219}, and \ce{WO3} \cite{Kumar2021, KIMURA2020} are widely used materials for hydrogen sensing due to their inherent semiconducting properties and compatibility with gas-sensing applications. 

However, in their pure, unmodified forms, these materials often exhibit limited sensing performance. In recent decades, efforts have been made to develop highly sensitive, fast-response materials for hydrogen detection. Several strategies for modifying the material have been employed. Two of the most used approaches are: i)~doping of the metal oxides with noble metals such as palladium (Pd), platinum (Pt) and gold (Au) \cite{Yifan2017, Zou2016, PANTA2018112}. ii)~Another widely used method is nanostructuring \cite{Jagdale2017, Kozhushner2013}. Reducing the grain/particle size makes them nanostructured, increases the surface area, and provides more surface area available for gas interaction. This leads to a higher density of active sites, which is crucial for improving the material's ability to detect low concentrations of hydrogen gas. 

Metal oxide hydrogen gas sensors can be prepared via various techniques such as sputtering \cite{Kumar2021}, spray pyrolysis \cite{ABOUHELAL2002}, thermal evaporation \cite{Yushu2021}, and sol-gel methods \cite{Bessekhouad2003}. In this study, the sputtering technique is used, exploiting the benefits of sputtering deposition, such as its ability to provide a cleaner, highly controlled, and reproducible environment during the deposition process. 

In this article, we explore the capabilities of a combination of two methods/strategies of nanostructuring: i)~Partial replacement of Argon (Ar) with Helium (He) to create more nanostructured (porous) films, and ii)~glancing angle deposition (GLAD), commonly used for creating nanostructures. As a material to be altered, we selected titania with small copper doping, in this work, acting just as an agent to improve the conductivity of the material. We focus solely on the nanostructuring effect on the response of the film. And we demonstrate the enhanced response of the \ce{Cu}:\ce{TiO2}) to hydrogen. This improvement of the sensing performance based on the structural tuning is described and discussed thoroughly. It is important to emphasise that the found improvements are achieved without using any noble metal. The mechanisms of nanostructured films formation are discussed in detail, as well as the comparison of both approaches.

Helium incorporation during reactive magnetron sputtering—a technique sparsely documented for metallic sputtering and exceptionally rare for metal oxide semiconductor (MOS) deposition—receives particular scrutiny due to its pivotal role in achieving the observed nanostructuring and sensing improvements.

\section{Experimental}

 \subsection{Synthesis}
A reactive DC (direct current) magnetron sputtering was used for deposition of Cu-doped \ce{TiO2} \hl{ films in a LH Z400 (Leybold Heraeus, Germany) cylindrical chamber (25 L volume) using}  pure titanium target (Ti, 99.99\%) with an adjacent piece of copper wire placed inside the race track was sputtered in a constant-power regime at 100\,W. The system was pumped by a turbomolecular pump backed up with a scroll pump, achieving a base pressure of 1\,mPa. The substrates were kept at room temperature (RT) during depositions. The substrates were positioned at: a) Normal Angle Deposition (NAD) at 90 degrees and b) Glancing Angle Deposition (GLAD) at 5 degrees relative to the target's normal.  The distance from the target to the substrate was 55\,mm. 

The working pressure was kept constant at a value as close as possible to 0.53 Pa for all depositions, while various mixtures of argon, oxygen, and helium were used.  The denomination of samples is based on the fraction of helium that replaces the argon partially in the working. More precisely, it is the ratio of the partial pressure of helium ($p_\mathrm{He}$) and the sum of the partial pressures of the inert gases ($p_\mathrm{He}+p_\mathrm{Ar}$). Further described experiments used four various fractions: 0\%, 74\%, 78\%, and 82\%, and samples are denominated accordingly. The fraction of oxygen varied for particular depositions and is specified in the Results section in Table \ref{tab:pressures}.

After the deposition, the films were annealed at 500\,°C in air for 4 hours in the furnace.

\hl{Various substrates were used for deposition to enable the analyses described below. The appearance and structure of the films varied only slightly across different substrates. For conciseness, we present results obtained on the most appropriate substrate in the main discussion. A comparison of SEM micrographs and XRD patterns for different substrates is provided in the Supplementary Data (Figs. S1 and S2, respectively).}

 \subsection{Compostion and morphology}

 Films deposited on $10\times10\,\mathrm{mm}^2$ silicon (100) substrates were used for structural and compositional characterisation techniques, including X-ray Diffraction (XRD), Scanning Electron Microscopy (SEM), and Energy/Wavelength Dispersive Spectroscopy (EDS/WDS). 

 X-ray diffraction (XRD) analyses were conducted using a diffractometer (X'Pert PRO, PANalytical, UK) employing a Cu\,K$_{\alpha}$  radiation source configured in Bragg–Brentano geometry. The films’ morphology was investigated via scanning electron microscopy (SEM) (SU-70, Hitachi Ltd., Japan), and copper concentration in the doped titania films was quantified on exclusively deposited 300-nm-thick films on the Si substrate via WDS (MagnaRay, Thermo Scientific) at a low primary electron energy of 5\,keV. Standard reference samples of pure Ti, Cu, Rutile (\ce{TiO2}) (Astimex Scientific Ltd.) were used, revealing 0.4\,at.\,\% Cu in both NAD and GLAD samples.

\subsection{Surface area assesment}
 
 To check the effect of He admixing in the working atmosphere on the films' porosity, we used two approaches. One is the electrochemical measurements of polarisation resistance used for relative comparison of surface active areas of various samples. The second is direct measurement of the films' areal mass density using a quartz coating monitor, Quartz Crystal Microbalance (QCM).

 \subsubsection{Polarisation resistance measurements}
    
    Films deposited on stainless steel (SS) substrate were used for electrochemical measurements. These were conducted in a water-jacketed, three-electrode glass corrosion cell containing 120\,mL of 0.1M~NaCl electrolyte solution at room temperature. The samples, serving as the working electrodes, were mounted on the PTFE cell wall and sealed with a Viton gasket, exposing a circular area of $0.35\,\mathrm{cm}^{2}$ to the electrolyte. An Ag/AgCl (3M~KCl) reference electrode was positioned near the working electrode, while a platinum wire functioned as the counter electrode. All electrodes were connected to a potentiostat (Squidstat Plus, Admiral Instruments, USA). Potentiodynamic polarisation measurements were conducted, ranging from $-0.8\,\mathrm{V}$ to $+0.8\,\mathrm{V}$ versus Ag/AgCl, and were recorded at a scan rate of $1\,\mathrm{mV\cdot{}s^{-1}}$.
    
    \newcommand{\Rp}{\ensuremath{R_\mathrm{p}}\xspace}
    \newcommand{\Ra}{\ensuremath{R_\mathrm{a}}\xspace}
    \newcommand{\Rg}{\ensuremath{R_\mathrm{g}}\xspace}
    
    The Electrochemical corrosion experiment was used to calculate the polarisation resistance (\Rp) using the Stern-Geary equation \eqref{eq:stern-geary} \cite{HUDAK2023111376, WENG1997147, Elsener1991}:
    \begin{equation}
    \Rp = \frac{\beta_a \cdot \beta_c}{2.303 \cdot i_{\text{c}} \cdot (\beta_a + \beta_c)}, \label{eq:stern-geary}
    \end{equation}%
    using cathodic and anodic slopes ($\beta_c$ and $\beta_a$), as well as corrosion current density values ($i_c$) derived from Tafel plot.  For further details follow the references. In the latter text we describe how the \Rp reflects the relative changes in surface area.
    
 \subsubsection{Areal mass density measurements}
 
    For this analysis, dedicated depositions were performed using the precise QCM system. The films had been deposited on silicon substrates placed in proximity to the QCM. QCM measures the area mass density, $m_\mathrm{a}$, of the thin deposited film in the units of micrograms per square centimetre ($\mathrm{\mu{}g\cdot{}cm^{-2}}$). The real thickness of the film, $t$, was derived both from cross-sectional SEM micrographs of broken films on Si and atomic force microscopy (AFM, AIST-NT SmartSPM, Horiba Inc., France) scans on the partially masked substrates. The density of the film was then determined simply as
    \begin{equation}
    \rho = \frac{m_\mathrm{a}}{t}.
    \label{eq:density_simple}
    \end{equation}
    
    Such calculated densities are compared among the samples. The porosity of samples with non-zero helium in the working gas mixture is then calculated as:
    
    \begin{equation}
    \text{porosity} = \left(\frac{\rho_0 - \rho_{\text{He}}}{\rho_0}\right) \times 100\,\%,
    \label{eq:porosity}
    \end{equation}
    where $\rho_\mathrm{He}$ denotes the density of a particular film, $\rho_0$ denotes the density of film deposited with 0\,\% He fraction.
    
    For these measurements, a different system needed to be used. The chamber made of DN200 ISO-K six-way cross piping was equipped with a planar circular magnetron with a variable magnetic field (VT100, Gencoa, UK) and a 4" Ti target. The chamber was evacuated by a turbo-molecular pump backed up with a scroll pump down to 0.5\,mPa. The magnetron was powered by a DC power supply (GS10, ADL, USA). The operating conditions were set to closely match the conditions during film depositions in the previously described Leybold Heraeus system. The total pressure was 0.5 Pa. The target power was 140 Watt, which corresponds to the same target power density in the other system, considering the larger target diameter.
 
\subsection{Sensing measurements}

    Films deposited on $10\times10\,\mathrm{mm}^2$ quartz glass substrates were utilised to check the sensing performance. For this, square-shaped platinum electrodes with dimensions $3\times3\,\mathrm{mm}^2$ were overdeposited on samples using a simple sputtering coater (Q150T, Quorum Technologies, UK). Sensing evaluations were conducted at fixed temperatures of 300\,°C in a 1\% fixed hydrogen concentration in synthetic air.
     
    The gas sensing capabilities of the samples were evaluated using a custom-designed four-point probe system; more details are available in \cite{Haviar2025}. A constant flow of dry air with varying hydrogen fraction was mixed via mass flow controllers. The electrical resistance of the samples stabilised after approximately two hours, as described in our previous publications \cite{Kumar2020, Nirmal2025}. For this study, the sensitivity, $S$, is defined as:
    \begin{equation}
    S = \frac{\Ra}{\Rg}, \label{eq:sensitivity}
    \end{equation}
    where \Ra and \Rg denote the resistances in synthetic air and hydrogen-enriched environment, respectively. This definition is consistent with other studies on $n$-type MOSs.
    
    In this paper, the basic model of the hydrogen sensing mechanism is considered, \latin{e.g.}, \cite{Nirmal2025}. In n-type MOS, oxygen molecules from the ambient environment are physisorbed at the surface. The adsorbed particles, Eq.~\ref{eq:adsorb}, capture electrons from the material, leading to the formation of various oxygen ion species, Eqs.~\ref{eq:o2minus}--\ref{eq:ominusminus}. The reaction can be described as follows:
    
\begin{equation}
O_2 (\text{gas}) \rightarrow O_2 (\text{ads}), \label{eq:adsorb}
\end{equation}
\begin{equation}
O_2 (\text{ads}) + e^- \rightarrow O_2^- (\text{ads}), \label{eq:o2minus}
\end{equation}
\begin{equation}
O_2^- (\text{ads}) + e^- \rightarrow 2O^- (\text{ads}), \label{eq:ominus}
\end{equation}
\begin{equation}
O^- (\text{ads}) + e^- \rightarrow O^{--} (\text{ads}). \label{eq:ominusminus}
\end{equation}%
Although different ion species are preferred for different temperatures, this level of detail is not important for this study and we froward reader to other \cite{Ciftyurek2023, Li2024, Shimanoe2009} 

The basic model further assumes, that when hydrogen is introduced, it reacts with the preadsorbed oxygen species, forming water vapour leading to the returning of electrons back to the material \cite{Sheng2026, Kumar202018066, Haviar201822756}. This can be described by following Eqs.~\ref{eq:watervapours1}--\ref{eq:watervapours3}:  

\begin{equation}
O_2^- (\text{ads}) + 2H_2 \rightarrow 2H_2O (\text{gas}) + e^-, \label{eq:watervapours1}
\end{equation}
\begin{equation}
O^- (\text{ads}) + H_2 \rightarrow H_2O (\text{gas}) + e^-, \label{eq:watervapours2}
\end{equation}
\begin{equation}
O^{--} (\text{ads}) + H_2 \rightarrow H_2O (\text{gas}) + 2e^-. \label{eq:watervapours3}
\end{equation}

\section{Results and Discussion}

An honest understanding of the reactive sputter deposition process is crucial for controlling the synthesis of the described nanostructured films. Therefore, the aforementioned results are preceded by a short subsection describing the properties of the reactive magnetron sputtering process used. A comment on the use of copper doping is also included. 

\subsection{Reactive sputtering conditions tuning}

\begin{table*}[]
\centering
\caption{Overview of pressure and flow ratios during the reactive sputtering deposition of studied thin films.}
\label{tab:pressures}
\begin{tabular}{cccccccccc}
\hline
Sample name                                                          & \multicolumn{3}{c}{Flows}   & Total pressure    & \multicolumn{4}{c}{Pressures – without discharge}                        & Deposition rate              \\ \cline{2-4} \cline{6-9}
\multirow{2}{*}{$\frac{p_\mathrm{He}}{p_\mathrm{He}+p_\mathrm{Ar}}$} & \ce{He} & \ce{Ar} & \ce{O2} & during deposition & $p_\mathrm{He}$ & $p_\mathrm{Ar}$ & $p_\mathrm{O2}$ & $p_\mathrm{total}$ & at NAD                       \\
                                                                     & \multicolumn{3}{c}{[sccm]}  & [mPa]             & \multicolumn{4}{c}{[mPa]}                                                & [$\mathrm{nm\cdot{}min^-1}$] \\ \hline
0\,\%                                                                & 0.0     & 15.0~   & 0.9     & 498               & 0               & 500             & 30              & 530                & 50                           \\
74\,\%                                                               & 25.0    & 3.6     & 0.6     & 492               & 380             & 135             & 20              & 535                & 38                           \\
78\,\%                                                               & 26.6    & 2.9     & 0.5     & 494               & 400             & 114             & 17              & 531                & 37                           \\
82\,\%                                                               & 28.1    & 2.2     & 0.4     & 489               & 423             & 95              & 13              & 531                & 36                           \\ \hline
\end{tabular}
\end{table*}

\begin{figure}[bt]
  \centering
  \includegraphics[width=85mm]{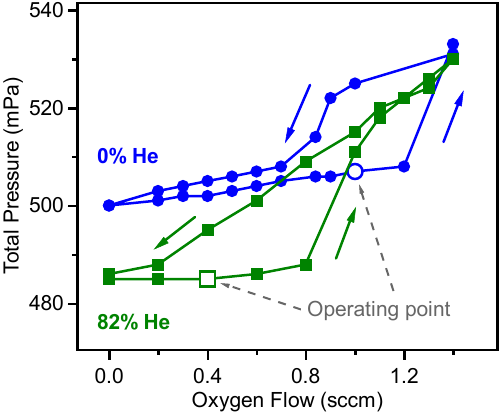}
 \caption{Hysteresis effect for segmented target of Ti and Cu during reactive sputter deposition for two selected He ratios: 0\%\,He sample (blue) and 82\%\,He sample  (green) working gas mixtures.}\label{fig:hysteresis}
\end{figure}

Pure titanium dioxide films synthesised by reactive magnetron sputtering in our system are highly insulating, featuring resistivities in the order of $10^6\,\mathrm{\Omega\cdot cm}$. Such high resistivities pose a significant challenge for conductometric gas sensing applications. To address this limitation, doping of titania with a sub-1\,at\% level of copper has been done to enhance the electrical conductivity of the material \cite{KARTHIK20106829}. Cu doping process, however, can be challenging due to the distinct behaviours of titanium and copper during reactive sputtering. Titanium segments of the target are easily poisoned, in contrast to the resilience of copper to that. This leads to a significantly higher sputtering yield of the copper segment, making a desired doping level difficult to achieve in the oxide sputtering mode. Deposition in a purely metallic mode followed by post-annealing is not a viable solution, as the strong oxidation of the films during post-annealing significantly alters the favourable microstructure formed during sputtering, which would contradict the purpose of this study. A functional approach that was utilised involves deposition in metallic mode, yet with the maximum possible amount of oxygen introduced into the working gas mixture, ensuring that the discharge does not switch into transition or oxide mode. This region was successfully identified by recording the hysteresis behaviour, as shown in Fig.~\ref{fig:hysteresis}, where the measured pressure vs oxygen flow curves are shown. Depending on the helium portion in the working gas mixture, the oxygen flow values were selected as 0.9, 0.6, 0.5, and 0.4\,sccm for 0\,\%, 74\,\%, 78\,\%, and 82\,\% He depositions, respectively. The lower acceptable amount of oxygen for the helium-rich mixture is due to the fact that the sputtering yield is much lower for helium ions than for argon ions. With the lack of argon ions in the helium-rich environment, the target is more prone to poisoning. 
The overview of the used mixtures of gases used together with the achieved deposition rate is summarised in Table \ref{tab:pressures}.

\vspace{1em}
In the subsequent sections, we systematically characterise the morphology, crystalline structure, and porosity of the synthesised films. This analysis is followed by a detailed discussion of helium's influence on film microstructure, supported by a simple plasma modelling. Finally, the sensing performance is evaluated, with particular emphasis on the effects of structural modifications.

\subsection{Morphology}

\begin{figure*}[tb]
  \centering
  \includegraphics{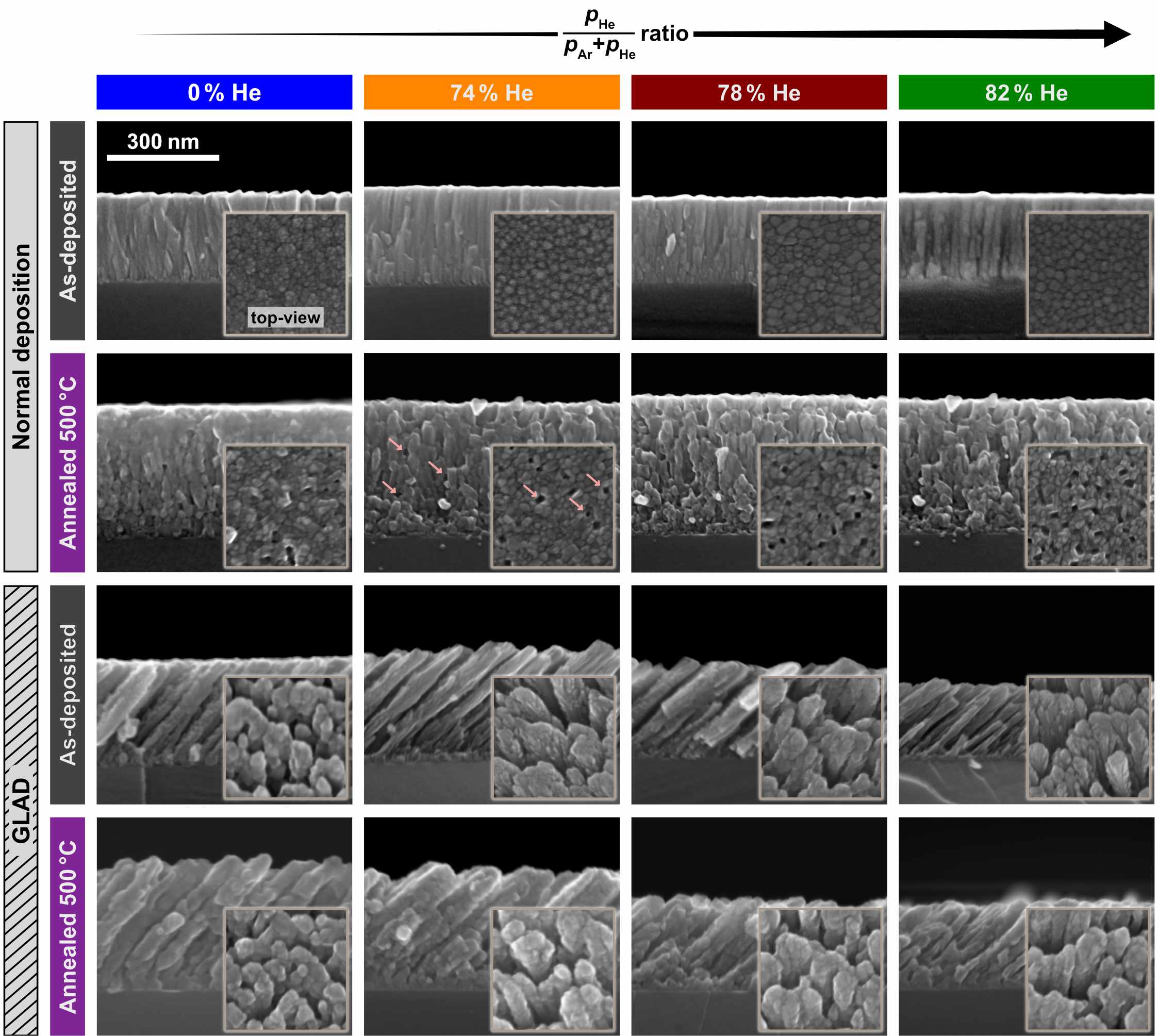} 
 \caption {SEM micrographs of NAD and GLAD samples under various conditions. All the top and cross-sectional views are the same scale indicated in the top left micrograph. Colour-coding of various helium portion in working gas is kept in the whole manuscript. \hl{The colour arrows point to example pores formed in the NAD films.} \hl{The comparison of micrographs of films deposited on various substrates can be found in the Supplementary Material, Fig.~S1.}}\label{fig:sem}
\end{figure*}

Fig.~\ref{fig:sem} shows the surface and cross-sectional micrographs of both as-deposited and annealed films.

SEM micrographs of the as-deposited NAD layers exhibit only subtle changes as the helium content in the working gas is varied. In the cross-sectional views, we observe slightly smaller apparent grains with increasing helium, while the top-view images appear all very similar, with only the 0\% He sample displaying a slightly finer structure atop the columns. The main differences, however, are observed in the annealed samples. In the NAD samples, there is a clear increase in the number of void structures, cavities, which are more numerous as the helium content increases. In the top-view images, it is evident that the structure becomes increasingly porous, with more open pores, \hl{indicated with colour arrows in the \mbox{Fig. \ref{fig:sem}}}, as the helium fraction in the working gas increases.

For the GLAD layers, the situation is different. The cross-sectional images of the as-deposited films reveal the emergence of a fine structure superimposed on the pronounced, characteristic tilted GLAD columnar structure. In the annealed samples, this effect is even more pronounced. Furthermore, the annealed columns in the GLAD structures are also more frequently interconnected and lack such sharp ends, as is most distinctly visible even for the 74\% He sample.

\subsection{Crystalline structure}
    
     \begin{figure*}[bt]
        \centering
        \includegraphics[width=190mm]{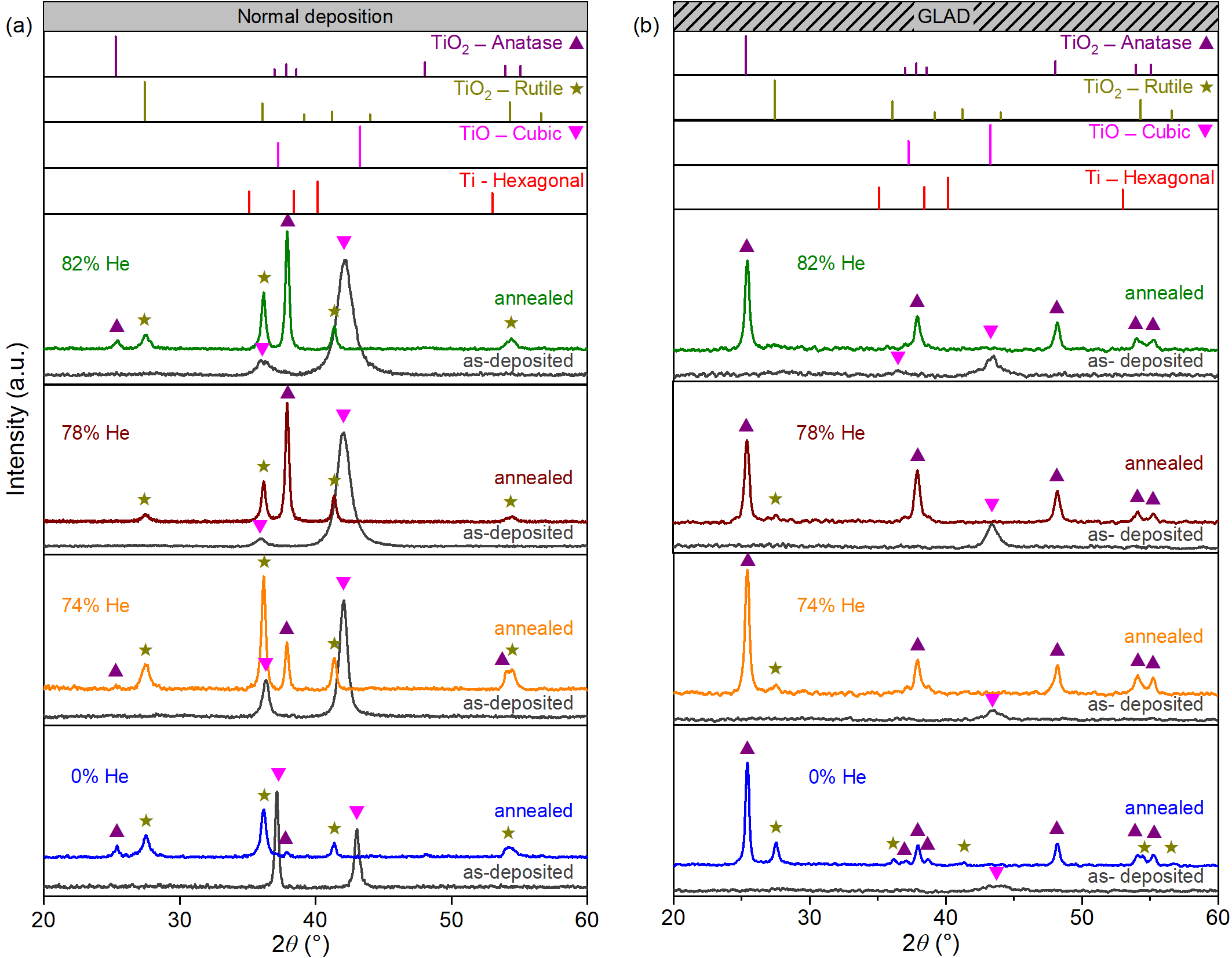}
        \caption{XRD spectrograms of (a) NAD and (b) GLAD films. As-deposited films are in grey, the annealed films are colour-coded. \hl{The comparison of spectrograms of films deposited on various substrates can be found in the Supplementary Material, Fig.~S2.}}
        \label{fig:xrd}
    \end{figure*}
    
    XRD spectrograms of NAD and GLAD films are shown in Fig. \ref{fig:xrd}a and \ref{fig:xrd}b, respectively. The spectrograms of as-deposited samples are depicted in grey for each helium fraction, while the annealed films' data are colour-coded.

    The NAD as-deposited films show a structure of sub-stoichiometric titanium oxide. Two distinct peaks at approximately 37° and 43° for 0\% He sample can be attributed to the (111) and (200) planes of a cubic TiO structure (PDF No. 01-073-8760, unit cell size of 4.177\,\AA) with a fitted unit cell size of 4.20\,\AA. The same cubic structure is identified for non-zero helium samples, but with an obvious shift of the peaks towards lower angles, better fitted by a larger cubic cell with a size of 4.28\,\AA. In addition, the peaks grow in width with the increasing helium fraction, indicating the shrinking of the crystallites. \hl{From Scherrer's equation, crystalite size can be estimated to decrease from 20\,nm to 5.5\,nm as helium fraction increases from 0\% to 82\%}.
    
    During the annealing, the structures are oxidised to \ce{TiO2} phases formed by a mixture of anatase (PDF No. 01-070-7348) and rutile (PDF No. 01-079-6029). The ratio of the anatase and rutile peaks varies with changing helium fraction, best notable for the peak at 37.9\textdegree, with a considerably higher portion of the anatase for 78\% and 82\% He samples. 


    The GLAD as-deposited samples exhibit a poorly crystalline structure of substoichiometric TiO, when only a broad peak of (200) at 43.2\textdegree{} is clearly visible. The annealed samples are all of anatase structure with a small addition of the rutile phase, more distinct for the 0\% He sample. This corresponds to the fact that the less dense anatase phase can readily form within the GLAD structure with isolated columns. \hl{In case of NAD samples, the helium creates bubbles, and these can effectively stabilise the anatase structure} \cite{KAJITA2021113420}. \hl{Generally, structures with larger surface favour the anatase phase, which has a lower surface energy than rutile} \cite{DASILVA2017103}. 

    No sign of metallic titanium is visible in the spectrograms (PDF No. 00-044-1294).

\subsection{Porosity and surface area assessment}

\newcommand{\gcmv}{$\mathrm{g\cdot{}cm^{-3}}$}
\newcommand{\gcma}{$\mathrm{\mu{}g\cdot{}cm^{-2}}$}
\newcommand{\Wcma}{$\Omega\cdot\mathrm{cm}^2$}
\renewcommand{\thefootnote}{\fnsymbol{footnote}}

\begin{table*}[b]\scriptsize
\caption{Measured thicknesses, mass densities, and calculated density change of films deposited under various conditions. For reactively sputter-deposited films, the polarisation resistances are shown for both as-deposited and annealed samples.}
\label{tab:porosity}
\begin{tabular}{cccccclcc} \cline{1-6} \cline{8-9}\\[-0.9em]
\multicolumn{2}{c}{Working gas  pressure portion} & & Cross-sectional    &\multicolumn{2}{c}{}                                                                                                        &  & \multicolumn{2}{c}{Polarization resistance}                                            \\\vspace{0.2em}
Helium            &   Oxygen    & \multirow{-2}{*}{QCM}                       & microscopy view & \multicolumn{2}{c}{\multirow{-2}{*}{Calculated}}                       &  & {\small\color[HTML]{424242} \textbf{As-deposited}} & {\small\color[HTML]{000000} \textbf{Annealed}} \\ \cline{1-6} \cline{8-9} \\[-0.9em]
[He]/[He+Ar]   &[O\li{2}]/[total]   & Mass                       & Thickness       & Density change\footnotemark[1]         & Porosity                &  & $R_\mathrm{p}$                               & $R_\mathrm{p}$                           \\
{[}\%]            &[\%]                &[\gcma]&[nm]       &[\gcmv]&                         &  &[\Wcma]             &[\Wcma]              \\[0.1em] \cline{1-6} \cline{8-9} \\[-0.9em]
0                 & 0                    & 143                        & $315\,\pm\,14$  & 4.54                       &                         &  &                                              &                                          \\
0                 & 7.0                  & 145                        & $325\,\pm\,14$  & 4.46                       & \multirow{-2}{*}{0\,\%\footnotemark[2]} &  & 4.45                                         & 18.3                                     \\
82                & 0                    & 160                        & $385\,\pm\,15$  & 4.16                       & 8\,\%                   &  &                                              &                                          \\
82                & 3.2                  & 161                        & $390\,\pm\,21$  & 4.13                       & 7\,\%                   &  & 2.98                                         & 4.92 \\ \hline                                 
\end{tabular}\vspace{1em}

\footnotemark[1]\footnotesize Densities used for calculations and discussion from Materials Project \cite{MaterialsProject}:
\\[0.2em]
{
\centering
\begin{tabular}{lllc}
\hline
\multirow{2}{*}{Material}  & \multirow{2}{*}{Lattice} & \multirow{2}{*}{Record No.} &  Density\\
& & & [\gcmv] \\ \hline
Titanium metal (Ti)     & Hexagonal  & mp-46    & 4.59\\ 
Rutile (TiO\li{2})      & Tetragonal & mp-2657  & 4.24\\ 
Anatase (TiO\li{2})     & Tetragonal & mp-390   & 3.86\\ 
Titanium monoxide (TiO) & Cubic      & mp-2664  & 5.38\\ \hline
\end{tabular}
}

\vspace{0.5em}
\footnotemark[2]\footnotesize Assuming a completely solid film.
\end{table*}

Accurate determination of the true surface area or porosity of the deposited films is challenging. Therefore, we employed polarisation resistance measurements as a qualitative indicator of relative differences in reactive surface area, assuming comparable electrochemical surface behaviour across samples. \hl{Under this assumption, the polarisation resistance is then inversely proportional to the reactive surface area, which is in contact with the electrolyte and therefore takes part in the charge transport from or to the electrolyte.}

The right-hand columns of Table~\ref{tab:porosity} compare the polarisation resistance values for film prepared in an Ar+\ce{O2} mixture (\HeMin) with the one prepared under helium-assisted conditions (\HeMax). The polarisation resistance decreases when comparing the \HeMin and \HeMax samples, for both as-deposited and annealed structures. A lowered polarisation resistance implies a higher active surface area, consistent with the more open microstructure of He-assisted films. For both sample types, annealing increases the polarisation resistance, which can be attributed to i) changes in surface chemical composition (oxidation) and ii) a reduction in active surface area due to structural smearing. In \HeMin films, the increase after annealing is more than 4 times, whereas in \HeMax films, the change is only about 1.6 times. The \HeMin layers are relatively compact, with a low active surface area close to the geometrical surface area, and exhibit only narrow intercolumnar pores, typical of sputter-deposited films. Upon annealing, the columns grow in volume, further constricting the minimum gaps. On the contrary, the He-assisted films are porous from the beginning, and, due to the formation of apparent voids (see SEM results), the decrease in surface area from smearing during annealing is not as distinct.

Increase of surface area can also be described in terms of rising porosity. The porosity can be estimated from measurements of the real density of the deposited films, as shown in the left-hand columns of Table~\ref{tab:porosity}. The measured mass densities reveal that the \hl{NAD} \HeMax films (both with and without oxygen) are lower than the tabulated values for solid materials by approximately 7, and 8\,\%, respectively. However, this reduction cannot be attributed solely to pore formation and the associated increase in surface area, since XRD indicates a clear change in the lattice parameter.

\subsection{Discussion on He-assisted film growth}

 The data presented show that the film structure is influenced by the presence of helium in the working gas mixture. Our observations are consistent with findings from other groups that investigated the deposition of mostly metallic films. In \cite{Han2017}, the authors examined the effect of helium on the growth of titanium films and its ability to retain hydrogen. \hl{More works considering the helium implantation can also be found in the field of nuclear fusion, where authors study helium-plasma-irradiation-induced modification of metal surfaces, like tungsten \mbox{\cite{KAJITA2020100828}}, or ceramics, like silicon carbide \mbox{\cite{KOYANAGI2026156746}}}.
 
 Reactive deposition of titanium specifically is a well-studied phenomenon, \latin{e.g.}, \cite{Mraz2011, Zeman2002}; however, helium-assisted reactive deposition of titania has not been reported in the literature. A brief mention of the influence of helium under reactive conditions can be found, \latin{e.g.}, \cite{Gryse2001}, which discusses improved de-poisoning in the presence of light helium ions for sputtering in Ar+\ce{N2}+\ce{He} mixture.

Obviously, the composition of the working gas is crucial for sputtering yield as well as for the ionisation of the sputtering gas itself. It is generally assumed that the sputtering yield of helium ions is lower than that of argon ions and that He is more difficult to ionise (24.6\,eV for ionisation of He \cite{BONCHIN199970, SandraLBonchin, C3JA50065A} compared to 15.8\,eV for Ar \cite{S.Corde}). This behaviour was also observed in our experiments, where the deposition rate decreased in He-containing working mixtures. In \cite{Ibrahim2021}, where Al films were prepared in gas mixtures of helium and argon, a pronounced decrease in deposition rate was observed for He ratios exceeding approximately 90\,\%, which is close to the He ratio used in this paper.

\begin{figure*}[bt]
    \centering
    \includegraphics[width=190mm]{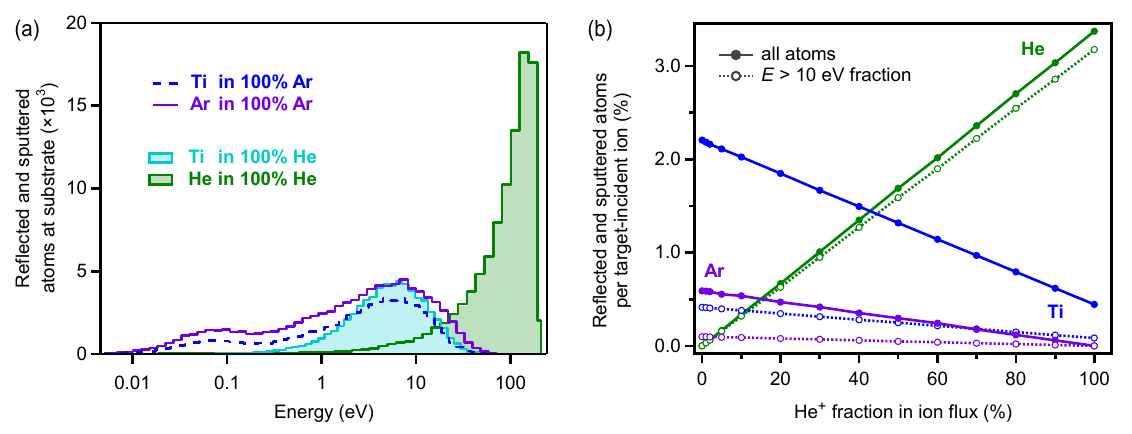}
    \caption{Results of SIMTRA simulation of selected particles arriving at substrate, the geometry is described in the experimental section. (a)~numbers of reflected Ar or He atoms and sputtered Ti atoms arriving at the substrate, considering $10^6$ emitted particles from the target. (b)~ratios of the above-mentioned atoms for various \ce{He+} fraction in the sputtering ion flux. Dashed lines with empty symbols represent the higher energetic part of the distributions.}
    \label{fig:simtra}
\end{figure*}

The most interesting is the influence of helium on the morphology of the deposited layers. In \cite{Caballero-Hernandez2015}, the authors demonstrated the deposition of porous silicon films by using helium for sputtering. Modifications of aluminium layer appearance are discussed in \cite{Ibrahim2021}, and similar trends were reported for \ce{Ti-H_x-He} films in \cite{Han2017}. A key role in film growth is played by backscattered particles impinging on the substrate surface. For the chamber geometry used in this work, modelling was carried out using the SIMTRA code \cite{Depla2012}. Due to code limitations, only Ar–He gas mixtures were simulated, with oxygen omitted. Nevertheless, even this simplified approach reveals the key difference in the deposition of titanium in mixtures of argon and helium. Fig.~\ref{fig:simtra}a shows the energy histograms of sputtered Ti atoms and backscattered Ar and He neutrals for two extreme working-gas compositions: 100\% Ar and 100\% He. An evident difference is that backscattered helium neutrals are much faster than argon ones, having the maxima of the histograms at 100\,eV and 9\,eV, respectively. Thus, He neutrals may be implanted deep into the growing film. Due to less effective scattering of titanium atoms in a helium environment, we also observe that all of the Ti atoms maintain the kinetic energy upon leaving the target (peak around 9\,eV), while in argon, a portion of the Ti atoms is scattered and slowed down (peak at lower energies around 0.09\,eV).

The situation for various working-gas compositions is summarised in Fig.~\ref{fig:simtra}b, which shows relative fluxes of sputtered Ti atoms and backscattered Ar and He neutrals. In addition, curves for fast particles with energy above 10\,eV are displayed. It is important to understand that the dependencies are functions of the fraction of \ce{He+} ions arriving at the target rather than the fraction of He in the working gas. The performed simulations do not allow direct determination of this ratio. However, a qualitative estimate of the ion composition in the plasma can be inferred from the film properties. In \cite{Caballero-Hernandez2015, Ibrahim2021}, the helium content in metallic films was measured using EELS (electron energy loss spectroscopy) and RBS (Rutherford backscattering spectroscopy). For metallic films prepared in He+Ar mixtures \cite{Ibrahim2021}, the He content reached several atomic percents within the metal matrix. Comparing Ti and He fluxes in Fig.~\ref{fig:simtra}b, we see that units of atomic percents of He reach the substrate at low \ce{He+}-fraction conditions, approximately for values around 10\% of \ce{He+} at the target, which can be assumed to be a realistic ratio considering the poorer ionisation of helium.

The presence of helium in the films is also indicated by XRD results. For NAD as-deposited films prepared with helium (Fig. \ref{fig:xrd}a), a noticeable shift of TiO diffraction peaks toward lower diffraction angles is observed. This shift indicates an expansion of the crystalline lattice, likely caused by interstitial implantation of helium. The $2\theta$ displacement relative to the XRD standard (or films prepared without helium) is 0.94\degree{} for the peak (2 0 0), corresponding to an increase of approximately 0.089\,\AA{} in the lattice parameter $a = 4.29\,\mathrm{\AA}$. The resulting volume change of the primitive cell equals then 4.8\,\AA$^3$, \hl{which can be attributed to the implanted helium}. The size He-occupied volume can be roughly estimated from \latin{ab-initio} study of sputtered materials in Ref. \cite{Houska2006}. \hl{Here, the authors calculate the interatomic distance for an implanted helium in a ceramic material. The estimated minimum radius of the pore occupied by one helium atom is 1.9\,\AA. The ratio of the volume of such a pore (29 \,\AA$^3$) and the volume change derived from XRD (4.8\,\AA$^3$) is approx. 17\,\%}, meaning the portion of unit cells occupied by interstitial He. Accounting for eight atoms per unit cell in cubic TiO, we obtain a rough estimate of the He concentration in the films of around \hl{2}\,at.\%. Even though the size of the He pores was calculated for different materials, we get reasonable values in units of atomic percent from the sputtering studies. The measurements by QCM under similar sputtering conditions revealed the difference of the density of the \HeMax and \HeMin films to be of a value approximately 8\,\% which points to  slightly lower He content.

Importantly, the implanted helium further modifies the layer structure during post-annealing. During annealing, helium atoms (initially trapped at interstitial sites, bound to defects or grain boundaries) begin to form He clusters that evolve into small bubbles along grain boundaries \hl{\mbox{\cite{Fernandez20251, ISHIYAMA199690, KAJITA2014438, KAJITA2020100828}}}. These are directly observable in SEM micrographs (see NAD images in Fig.~\ref{fig:sem}). However, such behaviour is not expected for GLAD films, where helium atoms can more easily escape completely from the material, both during annealing and even during deposition.

In addition to the dominant influence of implanted backscattered helium atoms, two other effects influencing film growth can be discussed. The first is the reduction of the hammering effect \cite{Houška2016}: helium, due to its significantly lower atomic mass compared to argon, exerts a much weaker hammering effect during deposition, enabling the growth of less densified films. The second effect, also discussed in \cite{Ibrahim2021}, is the cooling effect of helium. Higher helium concentrations in the working gas reduce adatom mobility on the substrate surface, thereby affecting film growth dynamics. Lower mobility results in the formation of more nanostructured or porous films.

\subsection{Sensing response and morphological discussion}
\begin{figure*}[bt]
    \centering
    \includegraphics[width=190mm]{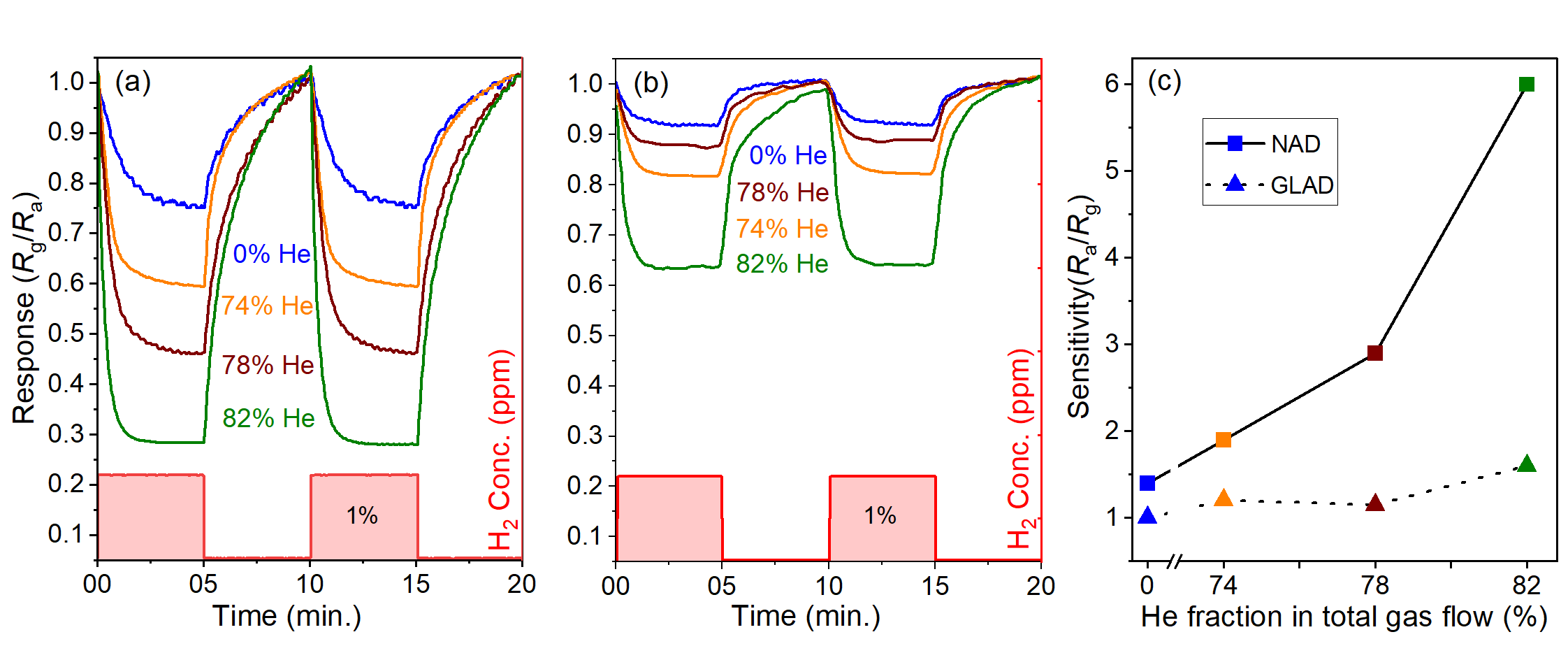}
    \caption{Normalised sensing response for (a) NAD films and (b) GLAD films. Sensitivity derived from the response by fitting (c).}
    \label{fig:response}
\end{figure*}

The normalised sensing response of NAD and GLAD films towards 1\,vol.\% \ce{H2} (in synthetic air) is plotted in Figs.~\ref{fig:response}a and \ref{fig:response}b, respectively, where the colour-coding of the helium fraction is kept consistent with the previous figures. The sensitivity values displayed in Fig.~\ref{fig:response}c were obtained by fitting an exponential decay to both the response and recovery parts of the response curve, and then by using Eq.~\ref{eq:sensitivity}.  

In both data representations (normalised response and sensitivity), it is visible that the GLAD films are less sensitive than the NAD films. Furthermore, introducing helium into the working gas mixture results in the formation of structures that perform better in terms of hydrogen sensitivity. For NAD films, this dependence is clearly monotonic. For GLAD samples, the difference is small for low helium fractions, and the GLAD~\HeMax\ sample shows approximately 50\% higher sensitivity than the GLAD~\HeMin\ one. The improvement induced by helium is far more pronounced for NAD samples, where the sensitivity increases from 1.4 to 6.0 as the helium fraction in the working gas rises from 0\,\% to 82\,\%.  

\begin{figure}[bt]
  \centering
  \includegraphics[width=85mm]{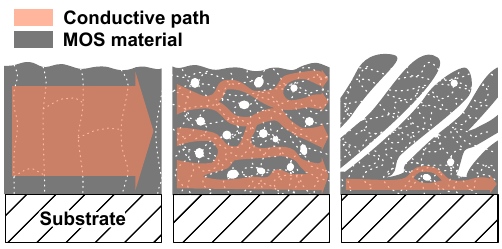}
 \caption{Cartoonistic comparison of conductive channel in densified thin film (\HeMin), porous thin film (\HeMax), and GLAD thin film.}\label{fig:cartoon}
\end{figure}

To explain this behaviour, we can look at the cartoonistic representation in Fig.~\ref{fig:cartoon}. Structures prepared with helium are porous (exhibit lower compactness) and therefore provide a larger reactive surface where pre-adsorbed oxygen can react with the detected hydrogen. At the same time, we can assume that the NAD helium-containing films are more crystalline (see Fig.~\ref{fig:xrd}a). With increasing helium fraction, an increase in the portion of the anatase crystalline phase is also observed, with dominating reflection (004). Such pronounced texture may play a crucial role in sensing improvement because the (004) plane facilitates much faster hydrogen dissociation due to the low activation energy reported in literature \cite{ZHOU201920606}. \hl{However since our films expose many facets to the air, we believe that the texture, does not play a crucial role here, such as in studies done on monocrystals. Another important aspect is the role of the crystalline structure, which changes from variously mixed rutile-anatase to purely anatase structure in the case of GLAD films. Anatase generally exhibits a stronger hydrogen response than rutile, and many authors reported that mixed–phase anatase–rutile films outperform either single-phase \mbox{\cite{Li2024, Enachi2015, Zakrzewska2017}}. In GLAD films with an 82\% helium fraction, the phase is purely anatase, yet the hydrogen response surpasses that of mixed‑phase films, indicating that the crystal phase alone does not govern the response.}

To conclude, the main reason for the improvement in response likely lies in the change of morphology and the enlarged ratio of reactive surface to volume. The change of effective conductive channel, as schematically indicated in Fig.~\ref{fig:cartoon}a and \ref{fig:cartoon}b.

The enhancement in GLAD films is not as strong as might be expected and as sometimes reported in the literature \cite{HAN201740, article, Jyothilal, HORPRATHUM2013685, KWAN2012164, WISITSOORAT2013795}. The most probable reason is the presence of a thin “seed" layer, \cite{Jensen2005}, sometimes named as a "wetting" layer, \cite{Hawkeye2007}, layer at the substrate interface, which (i) diminishes the contribution of the highly porous GLAD structure in the upper part of the film to the conductive path. This effect is sketched in Fig.~\ref{fig:cartoon}c. However, the change of conductive path shape alone cannot explain the decrease in sensitivity, since the surface-to-path ratio for the very thin wetting layers at the bottom should still be advantageous. It is therefore also important that the bottom wetting layer (ii) is less accessible for the diffusion of both the target gas and preadsorbing oxygen, which reduces the overall response of the device. 

\section{Conclusion}

This study investigates the effects of helium (He) in stable plasma conditions on the structural and response behaviour of thin films. Helium was partially substituted for argon (Ar) while maintaining constant overall working pressure and input power. The incorporation of helium into the plasma modifies the structural and morphological properties of the thin films by influencing the sputtering yield and the dynamics of the ejected atoms. These changes lead to the formation of nanostructures or porosity within the films, driven by mechanisms such as backscattering, cooling, and the hammering effect of helium, as elaborated in this study. The impact of helium was verified using X-ray diffraction (XRD) analysis and detailed surface and cross-sectional imaging via scanning electron microscopy (SEM). Furthermore, the sensing performance of the films was evaluated, showing a remarkable enhancement with increasing helium content. At the maximum helium fraction (82\%), the sensing response improved fourfold for the NAD sample and doubled for the GLAD samples. This enhancement can be attributed to increased surface active area and porosity, as confirmed by Quartz Crystal Microbalance (QCM) measurements and electrochemical analysis. These findings demonstrate that incorporating helium into plasma is an effective strategy to enhance nanostructuring and porosity, thereby significantly improving the functional performance of thin films.
This study employed the helium-assisted sputter deposition for synthesis of Cu doped titania. To our knowledge, this is the first time the reported usage of this synthesis method has been used to tune the nanostructure of the metal oxide for hydrogen sensing purposes. The mechanisms of morphological modification are described here, though further study of this phenomenon will need to employ advanced analyses capable of detecting the helium directly within the films. Nevertheless, this method could be beneficial for modifying the sensing MOS materials, improving their sensing performance.

\section*{Acknowledgement}
This work was supported by the project Quantum materials for applications in sustainable technologies (QM4ST), funded as project No. CZ.02.01.01/00/22\_008/0004572 by Programme Johannes Amos Comenius, call Excellent Research.
The authors acknowledge professor Jiří Houška for his valuable insights into the growth process discussion, professor Jiří Čapek for his necessary help with the deposition system, and dr. Radomír Čerstvý for XRD measurements and analyses. All acknowledged scientists are from the University of West Bohemia in Pilsen.

\section*{Declaration of generative AI and AI-assisted technologies in the writing process}
During the preparation of this work, the authors used ChatGPT to get style and grammar recommendations for the draft text. After using this service, the authors reviewed and edited the content as needed and take full responsibility for the content of the published article.

\bibliographystyle{elsarticle-num}
\bibliography{biblio.bib}

\end{document}